\documentclass[showpacs,prl,aps,superscriptaddress,twocolumn,floatfix]{revtex4-1}
\usepackage[english]{babel}
\usepackage{amsmath}
\usepackage{amssymb,mathrsfs}
\usepackage{graphicx}% Include figure files

\begin{document}

\title{Oblique breathers generated by a flow of two-component Bose-Einstein
condensate past a polarized obstacle}

\author{A.~M.~Kamchatnov}
\author{Y.~V.~Kartashov}
\affiliation{Institute of Spectroscopy,
  Russian Academy of Sciences, Troitsk, Moscow, 142190, Russia}

\begin{abstract}
We predict that oblique breathers can be generated by a flow of two-component
Bose-Einstein condensate past a polarized obstacle which attracts one
component of the condensate and repels the other one. The breather exists
if intra-species interaction constants differ from the inter-species interaction
constant and it corresponds to the nonlinear excitation of the so-called
polarization mode with domination of the relative motion of the components.
Approximate analytical theory is developed for the case
of small-amplitude breathers that is in reasonable agreement with the exact
numerical results.
\end{abstract}


\maketitle

{\it Introduction.}---Flow of Bose-Einstein condensates (BECs) past obstacles reveals a number of
different nonlinear excitations of the condensate. For example, in a one-component condensate with
the sound velocity $c_s$, whose value is determined by the background density, nonlinear interaction
constant and the mass of atoms, the flow is superfluid for velocities $V$ satisfying the condition
$M\equiv V/c_s<M_c$ ($M_c\approx0.37$ for the flow past a disk in 2D geometry). For greater Mach numbers
$M>M_c$ vortices are generated by the flow what
means loss of superfluidity (see \cite{fpm-92,wma-99,br-2000,rica-01,pnb-2005}). Another channel of dissipation opens,
according to the Landau criterion, when the flow velocity exceeds the sound velocity, that is
at $M=1$ (see, e.g., \cite{km-01,ap-04}). In this case, the interference of Bogoliubov waves generated by a supersonic
flow leads to formation of a so-called ``ship-wave'' pattern located
outside the Mach cone \cite{chcs-06,gegk-07}. Inside the Mach cone the vortex streets are generated
in the flow velocity interval $1<M<1.44$, but for velocities with $M>1.44$
very specific oblique dark solitons \cite{kld-98} are generated \cite{egk-06,kp-08,kk-11}.
They have been observed in experiments \cite{amo-11,grosso-11} with flows
of polariton condensates past obstacles. These nonlinear structures manifest themselves
as dips in the distributions of the condensate's density.

Creation of two-component atomic \cite{myatt-97,skurn-98} and spinor polariton \cite{kasprzak-06,balili-07}
condensates triggered extensive research activity, both theoretical and experimental, centered around
nonlinear properties of two-species superfluids (see review articles \cite{frantz-10,cc-13} and references therein).
One of new properties of such superfluids is a possibility of formation of topological excitations \cite{fsm-11,half-sol-12}.
Another specific feature of two-species superfluids is the existence of two modes of motion which can be called {\it
density} waves and {\it polarization} waves---in the density waves the two species move mainly in phase
with each other whereas in the polarization waves they move mainly in counter phase. In the linear limit
there exist, correspondingly, two types of sound waves which in problems like a description of wave
patterns generated by the flow past an obstacle define two Mach cones.
Consequently, two types of ship waves are generated by
the flow of a two-component condensate past an obstacle.
The existence of two different types of excitations in two-component condensate suggests the possibility of
two different types of oblique dark solitons that may be generated upon the interaction of condensate with a defect.
However, previously only one type of oblique dark solitons \cite{gladush-09} have been
observed in the flow past a non-polarized
obstacle, i.e. the obstacle whose potential acts equally on both species of the condensate. In this
case the potential disturbs both species ``in phase'', that is the symmetry of the potential coincides
with the symmetry of the density waves, and only such waves are excited by the flow. This
suggests that another, previously elusive, polarization mode can be generated by a {\it polarized} obstacle
which acts differently on different species of the condensate.

In this Letter we study the properties of wave
patterns generated by the two-dimensional flow of a two-component condensate past such a polarized obstacle
and demonstrate that in this case
a previously unknown type of excitations enter the scene---{\it oblique breather}.
One of the most distinctive features of such breathers is that in contrast to conventional dark-dark solitons
generated by non-polarized potentials, the density distributions in two components of the breather have different
shapes with deep out-of-phase modulation.
The properties of oblique breathers are studied numerically in the framework of a system of
two coupled Gross-Pitaevskii (GP) equations and they are explained analytically in the small-amplitude
limit by reducing the GP system to a mKdV equation for a weakly nonlinear polarization mode.

{\it The model.}---In the mean field theory the dynamics of two-component BEC is described by the system
of GP equations in standard non-dimensional form
\begin{equation}\label{eq1}
\begin{split}
i\frac{\partial \psi _{1}}{\partial t}& +\frac{1}{2}\Delta \psi
_{1}-(g_{11}\left\vert \psi _{1}\right\vert ^{2}+g_{12}\left\vert \psi
_{2}\right\vert ^{2})\psi _{1}=\sigma_1U(\mathbf{r})\psi _{1}, \\
i\frac{\partial \psi _{2}}{\partial t}& +\frac{1}{2}\Delta \psi
_{2}-(g_{12}\left\vert \psi _{1}\right\vert ^{2}+g_{22}\left\vert \psi
_{2}\right\vert ^{2})\psi _{2}=\sigma_2U(\mathbf{r})\psi _{2},
\end{split}
\end{equation}
where Laplacian $\Delta $ acts on two spatial coordinates $\mathbf{r}=(x,y)$. We
assume that particles in both species have the same mass and the potential
$\sigma_kU(\mathbf{r},t)$ of the obstacle is repulsive if $\sigma_k=1$
and attractive if $\sigma_k=-1$. In our numerical simulations it is modeled
by the form
$    U(\mathbf{r})=U_0\exp(-\mathbf{r}^2/a^2)$
with $U_0=1.0,\,a=2$. The nonlinear interaction constants $g_{ik}$ are supposed to be
positive.
For simplicity we assume that both components have in an undisturbed uniform state the
same densities $\rho_1=\rho_2=\rho_0/2$, where $\rho_1=|\psi_1|^2$, $\rho_2=|\psi_2|^2$,
and $\rho_0$ is an undisturbed total density $\rho=\rho_1+\rho_2$ at $|\mathbf{r}|\to\infty$.

Linearization of the system (\ref{eq1}) for slightly disturbed background state
yields the dispersion relations for the linear waves in the two-component condensate
(see, e.g., \cite{gladush-09})
$\omega _{d,p}^{2}=c_{d,p}^2k^2+\tfrac14k^4,$
where
\begin{equation}\label{eq4}
c_{d,p}=\frac{1}{2}\left[\rho_0\left( g_{11}+g_{22}\pm
\sqrt{(g_{11}-g_{22})^{2}+4g_{12}^{2}}\right)\right]^{1/2}
\end{equation}
are the velocities of the density ($c_d$, upper sign) and polarization
($c_p$, lower sign) waves,  while $\omega_{d,p}$ are the frequencies describing temporal
evolution of perturbations $\propto \exp(-i\omega_{d,p}t)$.
The presence of two different velocities leads to the existence of
two Mach cones defined by the relations
\begin{equation}\label{eq5}
\sin \chi_{d,p }=\frac{c_{d,p}}{V}\equiv \frac{1}{M_{d,p }},
\end{equation}
where $M_{d,p }=V/c_{d,p }$ are the corresponding Mach numbers;
$\chi_{d,p }$ are the angles between the direction of the flow
and the lines representing the {\it density} and {\it polarization} cones.

{\it Oblique breathers.}---In order to demonstrate the principal difference between wave patterns
generated by non-polarized ($\sigma_1=\sigma_2=1$) and polarized ($\sigma_1=-\sigma_2=1$) obstacles,
we have solved the system (\ref{eq1}) numerically for these two cases
using the input conditions $\psi_{1,2}=(\rho_0/2)^{1/2}\exp(iVx)$. Typical results are illustrated
in Fig.~1. For the non-polarized obstacle the density ship waves are located
outside the density Mach cone and a dark-dark soliton is located inside it.
Remarkably, the existence of the polarization Mach
cone is not manifested at all in the density distributions of both condensate
components---linear and nonlinear polarization waves are not excited by the
non-polarized obstacle. In sharp contrast, the polarized obstacle leads to much richer dynamics
and generates both density ship waves (outside the outer Mach cone) and polarization ship
waves (outside the inner Mach cone). The density waves oscillate in phase in both
components what increases the amplitude of oscillations in the total density, whereas
the counter-phase oscillations in the condensate components in the polarization
ship waves lead to cancelation of oscillations in the total density. The polarized
obstacle does not excite a usual dark-dark soliton, but instead a more complicated structure
is generated in the vicinity of the polarization Mach cone. We shall call this structure
an {\it oblique breather} since, as we shall see below, it can be related with
time-dependent breather solutions of the associated nonlinear evolution equations.
\begin{figure}[tb]
\begin{center}
\includegraphics[scale=0.74,clip]{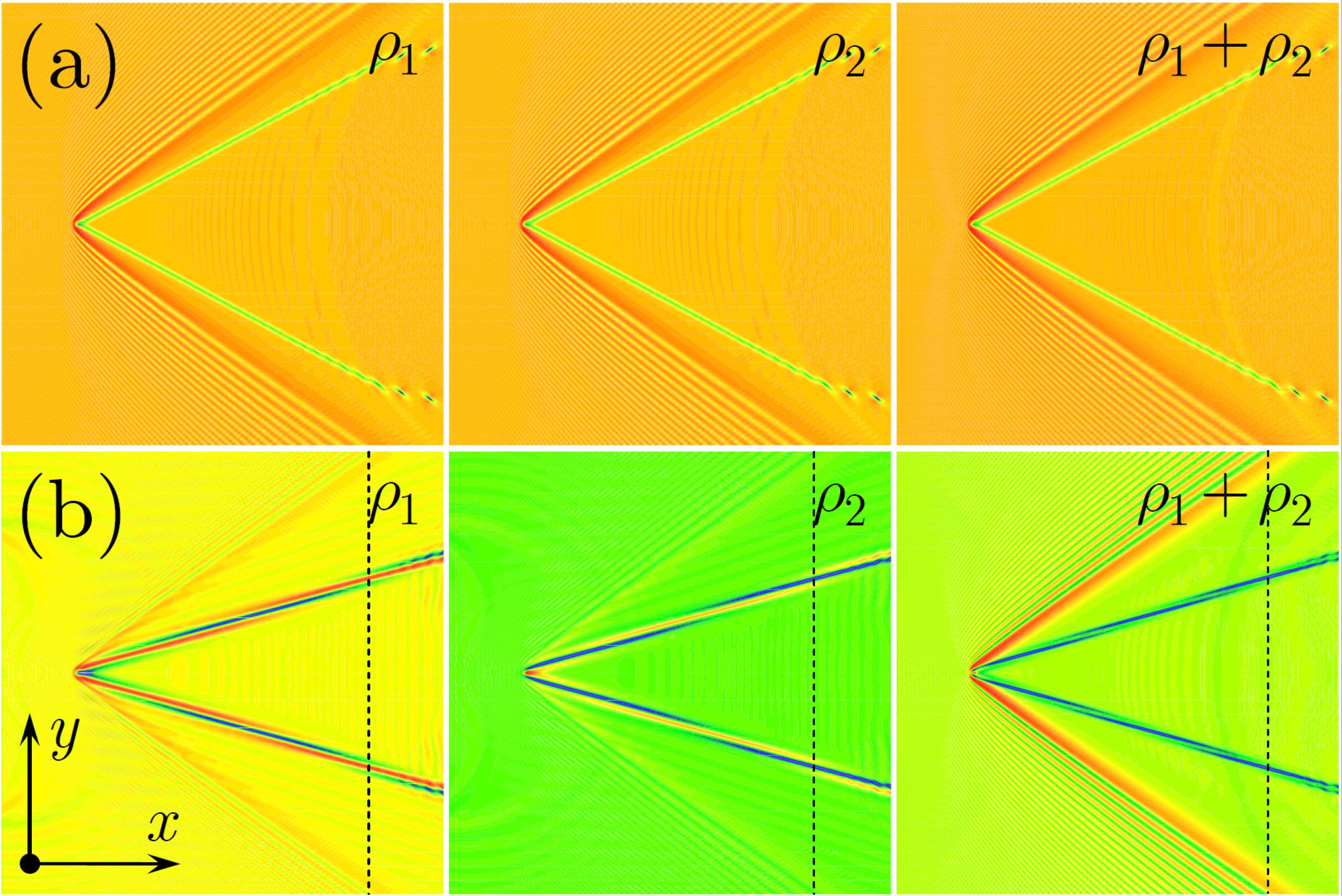}
\end{center}
\caption{(Color online.) Distributions of the densities of the first (left column) and second
(central column) components, as well as of total density (right column) for
(a) non-polarized obstacle with $\sigma_1=\sigma_2=1$,   and for (b) polarized obstacle with  $\sigma_1=1$,
$\sigma_2=-1$. In both cases  $g_{11}=g_{22}=1.0$, $g_{12}=0.6$, $V=2.3$, and $t=160$.}
\label{fig1}
\end{figure}

More detailed structure of the wave pattern generated by the polarized obstacle can be seen
in Fig.~2 representing the density distributions along $y$ axis at fixed value of
$x$ coordinate. The oblique
breather can be represented as a stationary spatially modulated nonlinear wave
with counter-phase nonlinear oscillations of the condensate components. When
its envelope is much narrower than the distance between two Mach cones, the space
between the oblique breather and the density ship waves is occupied by the
polarization ship waves clearly visible in Figs.~1 and 2.
\begin{figure}[tb]
\begin{center}
\includegraphics[scale=0.85,clip]{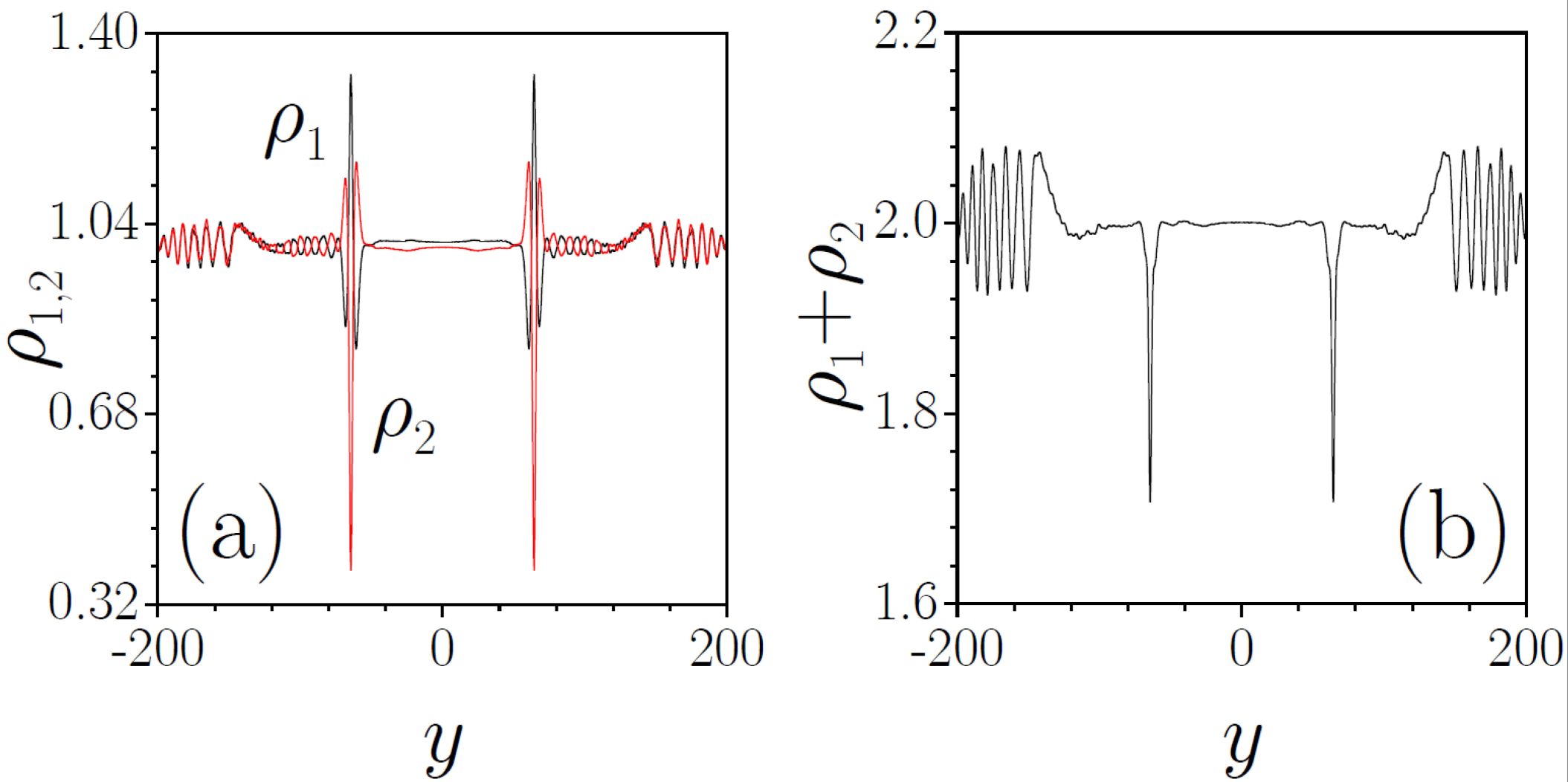}
\end{center}
\caption{(Color online.) The distributions of densities in two components (a) and total density (b)
along the $y$ axis at $x=100$, $t=160$  [these distributions correspond to dashed lines in Fig.~1(b)].
}
\label{fig2}
\end{figure}
The parameters of the oblique breather are determined by the
parameters of the incoming flow and those of the obstacle. It is worth noticing that the complete
cancelation of oscillations of the components in the total density distribution occurs only
if $g_{11}=g_{22}$, otherwise the linear eigenmodes correspond to a mixture of pure
density and polarization waves (see \cite{kklp-13}). At the same time, the wave pattern remains
qualitatively the same for small enough difference  $g_{11}-g_{22}$. The most important control
parameter is the incoming flow velocity $V$. In Fig.~3 we illustrate the modifications in the wave pattern
generated by the flow past a polarized obstacle with growth of $V$.
\begin{figure}[tb]
\begin{center}
\includegraphics[scale=0.74,clip]{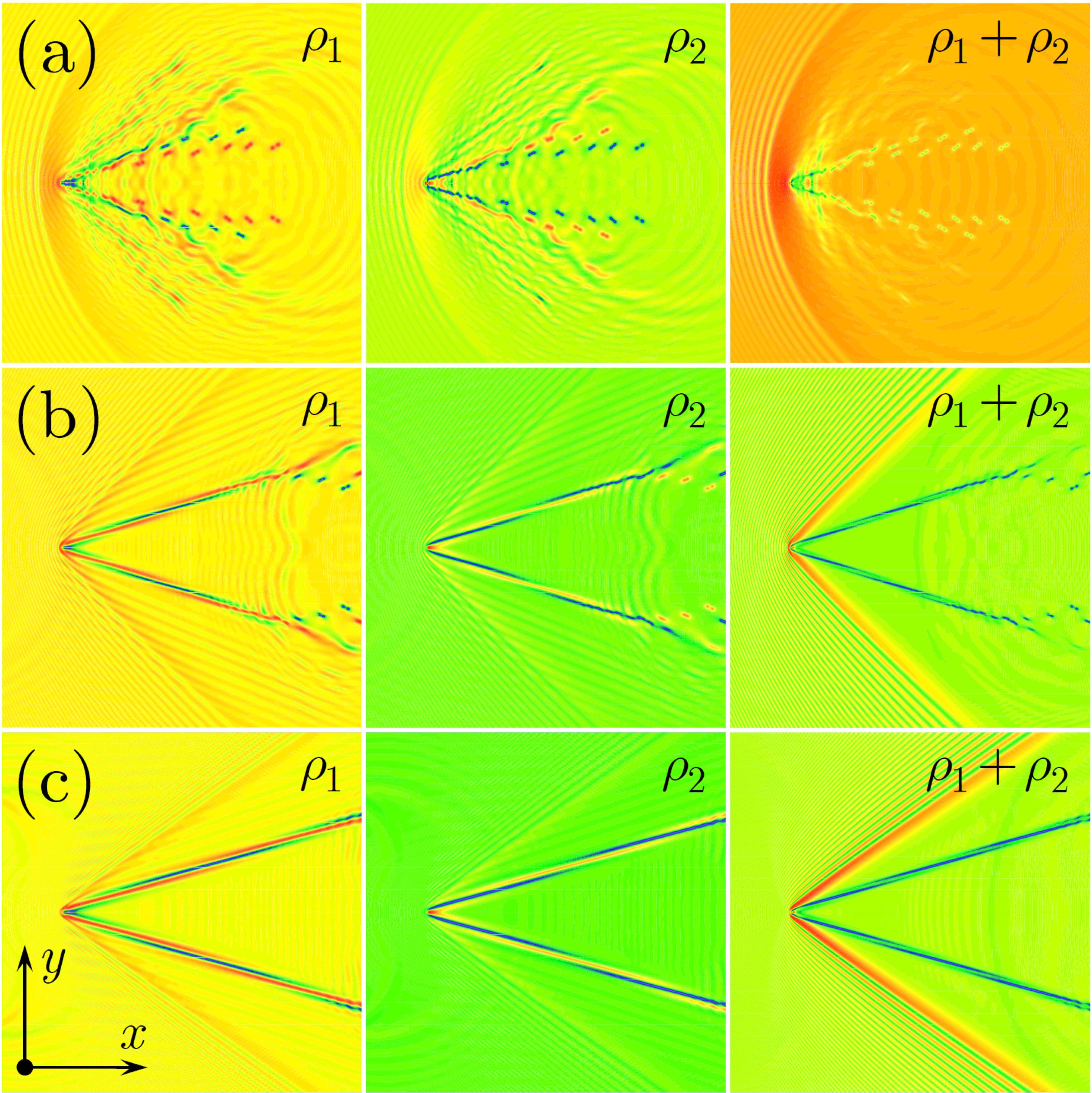}
\end{center}
\caption{(Color online.) Distributions of the densities of the first (left column) and second (central column)
components, as well as of total density (right column) at $t=160$  for the velocity of the flow $V=1.3$ (a),
$V=1.9$ (b), and $V=2.3$ (c). In all cases  $g_{11}=g_{22}=1.0$, $g_{12}=0.6$. }
\label{fig3}
\end{figure}
If $V=1.3$, then the breather is absolutely unstable and cannot be formed by the flow, and instead the
vortex streets are generated. If $V=1.9$, then a clearly resolvable breather is formed and further
increase of velocity to $V=2.3$ changes only its inclination angle with respect to the direction of the flow,
but does not change essentially its parameters. We suppose that this transition from a non-stationary vortex emission
to a stationary formation of oblique breather is physically similar to the transition from absolute instability of
oblique dark solitons to their convective instability studied in \cite{kp-08,kk-11,hi-12} for the one-component BEC flow.

{\it Analytical theory.}---Although the exact breather solutions of the coupled system of GP equations
(\ref{eq1}) are unknown, we develop here the approximate theory for the small-amplitude breathers which
explains with reasonable accuracy the observed features of new wave structures. To this end, we consider
a 1D version of the system (\ref{eq1}) which describes waves propagating along the $x'$ axis. For separation
of density and polarization modes it is convenient to introduce a spinor representation of the field
variables \cite{ktu-2005}
\begin{equation}\label{eq6}
    \left(
            \begin{array}{c}
              \psi_1 \\
              \psi_2 \\
            \end{array}
          \right)=
          \sqrt{\rho}e^{i\Phi/2}\chi=\sqrt{\rho}e^{i\Phi/2}
          \left(
            \begin{array}{c}
              \cos\frac{\theta}2\,e^{-i\phi/2} \\
              \sin\frac{\theta}2\,e^{i\phi/2}  \\
            \end{array}
          \right),
\end{equation}
where $\Phi$ has the meaning of the velocity potential of the in-phase motion; the angle $\theta$ is
the variable describing the relative density of
the two components ($\cos\theta=(\rho_1-\rho_2)/\rho$) and $\phi$ is the potential of the relative%
counter-phase motion. Correspondingly, the densities of the condensate components are given by
$    \rho_1=\rho\cos^2(\theta/2),\quad \rho_2=\rho\sin^2(\theta/2).$
In the uniform quiescent state of BEC with equal densities in the components we can take $\theta=\theta_0=\pi/4$ and then
the small-amplitude waves correspond to small variations of the relative density variable $\theta'\equiv\theta-\theta_0$
and small in-phase $U=\Phi_{x'}$ and counter-phase $v=\phi_{x'}$ velocities. The perturbation theory \cite{kklp-13} for polarization
waves with account of small dispersion and weak nonlinearity yields then the evolution equation which in the case
when $g_{11}=g_{22}\neq g_{12}$ has the form of mKdV equation for $\theta'$
\begin{equation}\label{eq8}
 \theta'_{t}+c_p\theta'_{x'}-\frac{3c_p(9g_{11}-g_{12})}{8g_{12}}
\theta^{\prime2}\theta'_{x'}-\frac1{8c_p}\theta'_{{x'}{x'}{x'}}=0.
\end{equation}
If its solution is found, then the other field variables are expressed in terms of $\theta'$
by the formulas
\begin{equation}\label{eq9}
     \rho=\rho_0-\frac{3c_p^2}{2g_{12}}\theta^{\prime2},\quad
    U=-\frac{c_p(3g_{11}+g_{12})}{2g_{12}}\theta^{\prime2},\quad
    v=2 c_p\theta'.
\end{equation}
The mKdV equation (\ref{eq8}) has a variety of solutions, among which there is the one-dimensional breather solution,
first presented in \cite{wadati-73}, which in our notations can be written as
\begin{equation}\label{eq10}
    \theta'=-\frac2{c_p}\sqrt{\frac{2g_{12}}{9g_{11}-g_{12}}}\frac{\partial}{\partial{x'}}
    \arctan\left(\frac{\eta\cos(\Theta_1+\beta_1)}{\xi\cosh(\Theta_2+\beta_2)}\right),
\end{equation}
where
\begin{equation}\label{eq11}
\begin{split}
   &\Theta_1=2\xi({x'}-c_pt)-\xi(\xi^2-3\eta^2)t/c_p,\\
   &\Theta_2=2\eta({x'}-c_pt)-\eta(3\xi^2-\eta^2)t/c_p,\\
   &\xi=\kappa\cos q,\quad \eta=\kappa\sin q,\\
   &\beta_1=p-q,\quad \beta_2=\ln(2\kappa\tan q/a)
   \end{split}
\end{equation}
and $p,\,q,\,\kappa,\,a$ are free parameters. This solution gives an approximate description
of the oblique breather pattern found above numerically when it is transformed to the appropriate reference frame
and the parameters are chosen in a proper way. The solution (\ref{eq10}) is written in the reference frame
associated with a quiescent condensate, where the breather propagates with the envelope velocity $V_b=c_p+(3\xi^2-\eta^2)/(2c_p)$
and the carrier wave velocity $V_c=c_p+(\xi^2-3\eta^2)/(2c_p)$ along
the ${x'}$ axis with the breather location line parallel to the ${y'}$ axis. We must transform it to the reference frame
with the obstacle located at the axes origin, the condensate's flow velocity directed along the $x$ axis,
and the breather location line inclined with respect to the $x$ axis at some angle $\chi_b$ chosen in such a way
that the breather becomes a stationary structure in the new reference frame. The same transformation
must compensate both the envelope velocity and the carrier wave velocity, $V_b=V_c$,  what gives $q=\pi/4$.
Besides that, these two velocities must be compensated by the projection of the flow velocity $V\sin\chi_b$ on the ${x'}$ axis.
That can be realized only at $c_p+\kappa^2/(2c_p)=V\sin\chi_b$. After these transformations the
phases $\Theta_1$ and $\Theta_2$ must be replaced by $\Theta_1=\Theta_2=\sqrt{2}\kappa\cos\chi_b(y\pm\tan\chi_b\cdot x)$
and, as a result, the solution (\ref{eq10}) transforms into the distribution of $\theta'$ in the $(x,y)$-plane.
Substitution of $\theta'(x,y)$ into Eqs.~(\ref{eq11}) yields the distributions of the other field variables.

It is convenient to express the parameter $\kappa$
in terms of the maximal amplitude of the density envelope $\Delta\rho\equiv|\rho-\rho_0|=24\kappa^2/(9g_{11}-g_{12})$.
As a result, we get, with account of the expression $c_p^2=\rho_0(g_{11}-g_{12})/2$ which is valid if $g_{11}=g_{22}$, a useful
relation between the breather angle $\chi_b$ and the maximal envelope amplitude $\Delta\rho$:
\begin{equation}\label{eq12}
    \sin\chi_b=\frac1{M_p}\left(1+\frac{9g_{11}-g_{12}}{24(g_{11}-g_{12})}\cdot\frac{\Delta\rho}{\rho_0}\right).
\end{equation}
Importantly, this formula predicts that the oblique breathers are located outside the polarization
Mach cone defined by Eq.~(\ref{eq5}).
Another important relation is given by the dependence of the inverse width $w=4\eta$
of the breather on its amplitude $\Delta\rho$:
\begin{equation}\label{eq13}
    w=\sqrt{(9g_{11}-g_{12})\Delta\rho/3}.
\end{equation}
\begin{figure}[tb]
\begin{center}
\includegraphics[scale=0.7,clip]{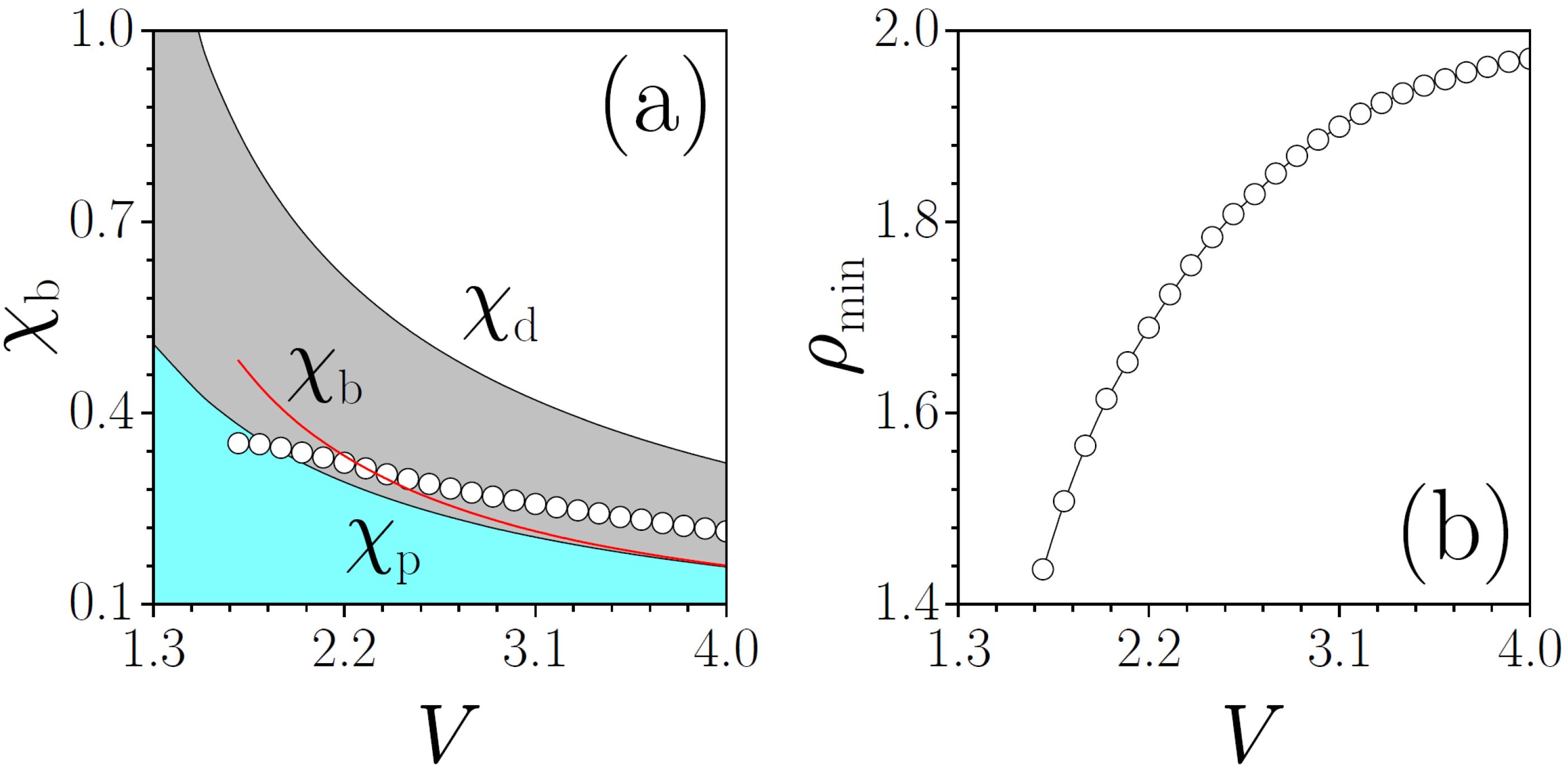}
\end{center}
\caption{(Color online.) The inclination angle $\chi$ (a) and depth $\rho_{min}=\rho_0-\Delta\rho$ (b) of the oblique breather
generated by the flow of condensate past the polarized obstacle versus flow velocity $V$ at $g_{11}=g_{22}=1.0$, $g_{12}=0.6$.
In (a) the line with circles shows numerical results, while red solid line shows analytical prediction
(\ref{eq12}) for the inclination angle. Blue region corresponds to the inner Max cone, while the region between
inner and outer Max cones is shown grey.
}
\label{fig4}
\end{figure}

Numerically found dependence of the breather angle $\chi_b$ as a function of $V$ is shown in Fig.~4(a) by circles.
As one can see, it agrees with qualitative prediction that oblique breathers are located outside the polarization Mach cone.
For comparison with the analytical formula (\ref{eq12}) we determined numerically
the minimal density $\rho_{min}=\rho_0-\Delta\rho$ in the breather as a function of $V$ which is plotted in Fig.~4(b).
Substitution of corresponding values of $\Delta\rho$ into Eq.~(\ref{eq12}) yields the dependence of $\chi_b$ on $V$ which
is shown in Fig.~4(a) by a red line. It perfectly agrees with numerical values of $\chi_b$ at velocities close to $V\approx2$,
however, the theoretical dependence deviates from the numerical one for $V\gtrsim3$. We explain this disagreement by the fact
that at $V\simeq3$ the breather's width given by Eq.~(\ref{eq13}) is of the same order as the distance between the two Mach cones and,
consequently, the breather cannot be considered as a structure well separated from other excitations---in this case the
interaction of the breather with the ship wave leads to deviations of the breather's parameters from
those calculated in the frames of perturbation theory. Another consequence of the interaction of the breather with the ship wave,
slowly changing along the Mach cone, is the finite lengths of wavecrests in the carrier wave.
\begin{figure}[tb]
\begin{center}
\includegraphics[scale=0.74,clip]{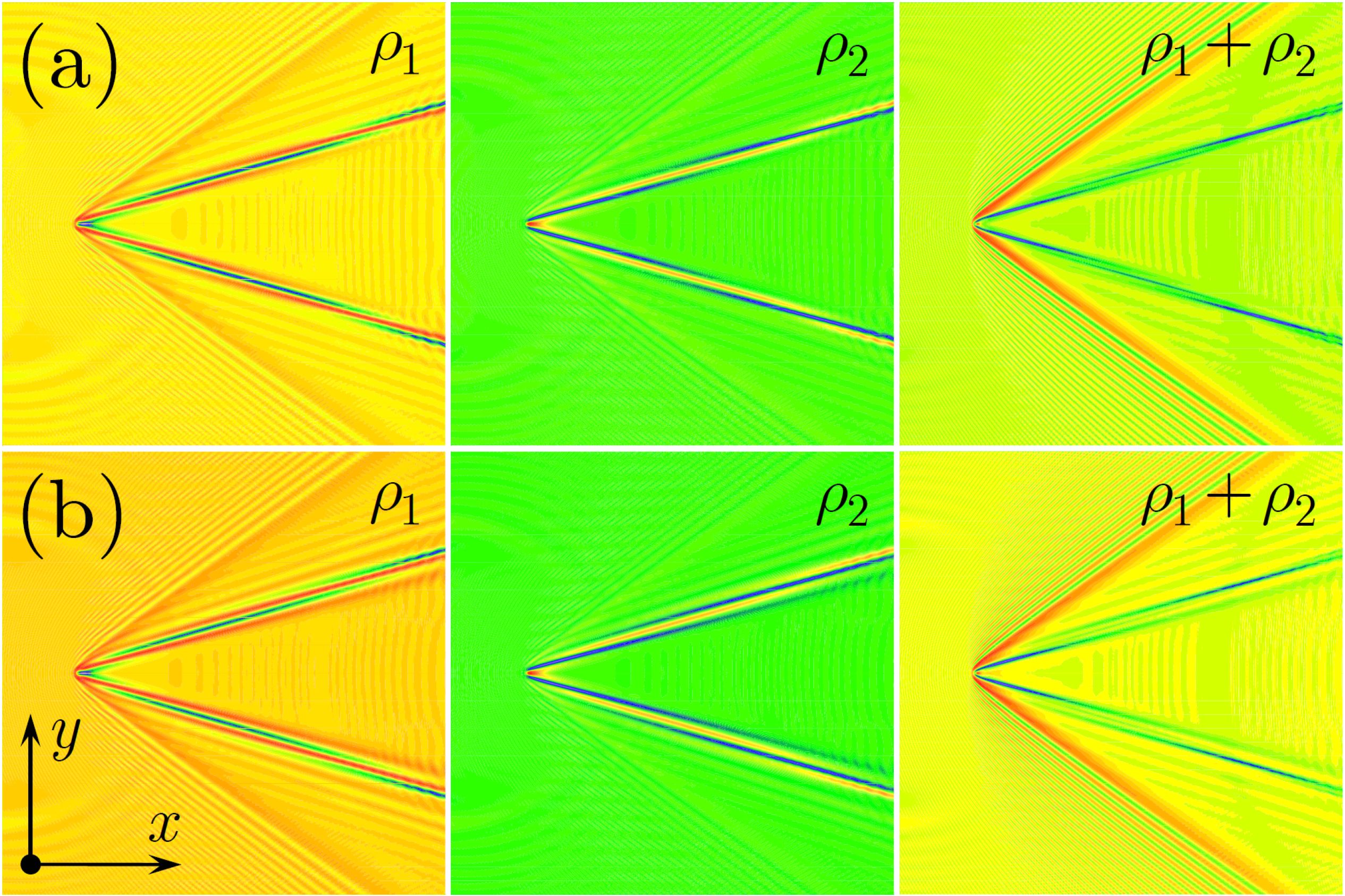}
\end{center}
\caption{(Color online.) Density distributions generated by the polarized obstacle at (a) $g_{11}=1.1$, $g_{22}=0.9$ and
(b) $g_{11}=1.2$, $g_{22}=0.8$. In both cases $V=2.3$, $t=160$, and $g_{12}=0.6$.
}
\label{fig5}
\end{figure}

{\it Arbitrary nonlinearity constants.}---The above theory was developed for the case when $g_{11}=g_{22}$.
Nevertheless, it is of considerable interest to study what happens if $g_{11}\neq g_{22}$.
Corresponding results obtained by direct solution of Eqs.~(\ref{eq1}) are presented in Fig.~5.
One can see that
now the counter-phase oscillations of the densities of components do not compensate each other and the ship wave pattern
becomes clearly visible in the distribution of the total density. With increase of the difference $g_{11}-g_{22}$ the width
of the oblique breathers also increases but qualitatively the whole wave pattern remains the same.

Summarizing, we have predicted that new nonlinear structures---oblique breathers---can be generated by
a flow of two-component condensate past polarized obstacles.


\begin{thebibliography}{99}

\bibitem{fpm-92} T. Frisch, Y. Pomeau, and S. Rica, Phys. Rev. Lett. {\bf 69,} 1644 (1992).

\bibitem{wma-99} T. Winiecki, J. F. McCann, and C. S. Adams, Phys. Rev. Lett. {\bf 82,}  5186 (1999).

\bibitem{br-2000} N. G. Berloff and P. H. Roberts, J. Phys. A: Math. Gen. {\bf 33,} 4025 (2000).

\bibitem{rica-01} S. Rica, Physica D {\bf 148,} 221 (2001).

\bibitem{pnb-2005} C.-T. Pham, C. Nore, and M.-\'E. Brachet, Physica D {\bf 210,} 203 (2005).

\bibitem{km-01} D. L. Kovrizhin and L. A. Maksimov, Phys. Lett. A {\bf 282,} 421 (2001).

\bibitem{ap-04} G. E. Astrakharchik and L. P. Pitaevskii, Phys. Rev. A {\bf 70,} 013608 (2004).

\bibitem{chcs-06} I. Carusotto, S. X. Hu, L. A. Collins, and A. Smerzi, Phys. Rev. Lett. {\bf 97,} 260403 (2006).

\bibitem{gegk-07} Yu. G. Gladush, G. A. El, A. Gammal, and A. M. Kamchatnov, Phys. Rev. A {\bf 75,} 033619 (2007).

\bibitem{kld-98} Y. S. Kivshar and B. Luther-Davies, Phys. Rep. {\bf 298,} 81 (1998).

\bibitem{egk-06} G. A. El, A. Gammal, and A. M. Kamchatnov, Phys. Rev. Lett. {\bf 97,} 180405 (2006).

\bibitem{kp-08} A. M. Kamchatnov and L. P. Pitaevskii, Phys. Rev. Lett. {\bf 100,} 160402 (2008).

\bibitem{kk-11} A. M. Kamchatnov and S. V. Korneev, Phys. Lett. {\bf 375,} 2577 (2011).

\bibitem{amo-11} A. Amo, S. Pigeon, D. Sunvitto, V. G. Sala, R. Hivet, I. Carusotto, F. Pisanello, G. Lem\'enager, R. Houdr\'e,
E. Giacobino, C. Ciuti, and A. Bramati, Science {\bf 332,} 1167 (2011).

\bibitem{grosso-11} G. Grosso, G. Nardin, F. Morier-Genoud, Y. L\'eger,
and B. Deveaud-Pl\'edran, Phys. Rev. Lett. {\bf 107,} 245301 (2011).

\bibitem{myatt-97} C. J. Myatt, E. A. Burt, R. W. Grist, E. A. Cornell, and C. E. Wiemann, Phys. Rev. Lett. {\bf 78,} 586 (1997).

\bibitem{skurn-98} D. M. Stamper-Kurn, M. R. Andrews, A. P. Chikkatur, S. Inouye,  H.-J. Miesner, J. Stenger,
and W. Ketterle, Phys. Rev. Lett., {\bf 80,} 2027 (1998).

\bibitem{kasprzak-06} J. Kasprzak, M. Richard, S. Kundermann, A. Baas, P. Jeambrun, J. M. J. Keeling, F. M. Marchetti,
M. H. Szymanska, R. Andr\'e, J. L. Staehli, V. Savona, P. B. Littlewood, B. Deveaud, and L. S. Dang, Nature, {\bf 443,} 409 (2006).

\bibitem{balili-07} R. Balili, V. Hartwell, D. Snoke, L. Pfeiffer, and K. West, Science, {\bf 316,} 5827 (2007).

\bibitem{frantz-10} D. J. Frantzeskakis, J. Phys. A: Math. Theor. {\bf 43,} 213001 (2010).

\bibitem{cc-13} I. Carusotto and C. Ciuti, Rev. Mod. Phys. {\bf 85,} 299 (2013).

\bibitem{fsm-11} H. Flayac, D. D. Solnyshkov, and G. Malpuech, Phys. Rev. B {\bf 83,} 193305 (2011).

\bibitem{half-sol-12} R. Hivet, H. Flayac, D. D. Solnyshkov, D. Tanese, T. Boulier, D. Andreoli, E. Giacobino, J. Bloch,
A. Bramati, G. Malpuech, and A. Amo, Nature Phys. {\bf 8,} 724 (2012).

\bibitem{gladush-09} Yu. G. Gladush, A. M. Kamchatnov, Z. Shi, P. G. Kevrekidis, D. J. Frantzeskakis, and B. A. Malomed,
Phys. Rev. A {\bf 79,} 033623 (2009).

\bibitem{kklp-13} A. M. Kamchatnov, Y. V. Kartashov, P.-\'E. Larr\'e, and N. Pavloff,   arXiv:1308.0784 (2013).

\bibitem{hi-12} M. A. Hoefer and B. Ilan, Multiscale Model. Simul., {\bf 10,} 306 (2012).

\bibitem{ktu-2005} K. Kasamatsu, M. Tsubota, and M. Ueda, Phys. Rev. A, {\bf 71,} 043611 (2005).

\bibitem{wadati-73} M. Wadati, J. Phys. Soc. Japan, {\bf 34,} 1289 (1973).


\end{thebibliography}
\end{document}